\documentclass[showpacs,twocolumn,prl]{revtex4}
\usepackage{amsmath}
\usepackage{amssymb}
\usepackage{color}
\usepackage{graphics}
\usepackage{epsfig}
\usepackage{floatflt}
\usepackage{hyperref}
\usepackage{epsfig}

\begin{document}

\title {Voltage-driven quantum oscillations in graphene}

\author{V.~A.~Yampol'skii}
\affiliation{Frontier Research System, The Institute of Physical
and Chemical Research (RIKEN), Wako-shi, Saitama, 351-0198, Japan}
\affiliation{A. Ya. Usikov Institute for Radiophysics and
Electronics National Academy of Sciences of Ukraine, 61085
Kharkov, Ukraine}
\author{Sergey Savel'ev}
\affiliation{Frontier Research System, The Institute of Physical
and Chemical Research (RIKEN), Wako-shi, Saitama, 351-0198, Japan}
\affiliation{Department of Physics, Loughborough University,
Loughborough LE11 3TU, UK}
\author{Franco Nori}
\affiliation{Frontier Research System, The Institute of Physical
and Chemical Research (RIKEN), Wako-shi, Saitama, 351-0198, Japan}
\affiliation{Department of Physics, Center for Theoretical
Physics, Applied Physics Program, Center for the Study of Complex
Systems, University of Michigan, Ann Arbor, MI 48109-1040, USA}

\date{\today}
\begin{abstract}
{We predict unusual (for non-relativistic quantum mechanics)
electron states in graphene, which are \emph{localized within} a
finite-width potential barrier. The density of localized states in
the sufficiently high and/or wide graphene barrier exhibits a
number of singularities at certain values of the energy. Such
singularities provide \emph{quantum oscillations} of both the
transport (e.g., conductivity) and thermodynamic properties of
graphene --- when increasing the barrier height and/or width,
similarly to the well-known Shubnikov-de-Haas (SdH) oscillations
of conductivity in pure metals. However, here the SdH-like
oscillations are driven by an \emph{electric} field instead of the
usual magnetically-driven SdH-oscillations. }
\end{abstract}  \maketitle

The Shubnikov-de-Haas effect, i.e., the oscillations of the
magneto-resistance of metals when increasing an external magnetic
field, was one of the first
macroscopic manifestations of the quantum-mechanical nature of
matter. The key to understanding this remarkable phenomenon was
pointed out by Landau
and Onsager
and it is described in many textbooks on solid state physics (see,
e.g., Ref.~\onlinecite{book}). Namely, electrons in the conduction
band of a metal in a strong magnetic field behave like simple
harmonic oscillators. The resulting energy spectrum is made up of
equidistant Landau levels separated by the cyclotron energy. The
density of electron states has singularities at the Landau levels.
When the magnetic field is changed, the positions of the Landau
levels move and pass periodically through the Fermi energy. As a
result, the population of electrons at the Fermi surface also
oscillates and, in turn, leads to quantum oscillations of the
transport and thermodynamic properties of a metal. The quantum
oscillations also manifest themselves in the thermoconductivity,
magnetization, sound attenuation, magnetostriction, and other
quantities.

These quantum oscillations are pronounced in conductors with a
long mean free path of charge carriers. This can occur in pure
metals, semimetals, and narrow band-gap semiconductors at low
temperatures, as well as in graphene, a one-atom-thick sheet of
carbon. The Shubnikov-de-Haas oscillations of the
magneto-resistivity were observed in graphene~\cite{osc1,osc2}
soon after its discovery~\cite{graph}. Due to the monolayer
honeycomb-lattice structure of graphene, its electrons obey a
massless Dirac-like equation (see, e.g.,
Refs.~\onlinecite{KP-graph,been}). This is responsible for the
unusual properties of graphene. In particular, the Landau levels
in graphene are not equidistant and these influence the period of
the Shubnikov-de-Haas oscillations~\cite{osc3,mur1}. Graphene has
another striking property: it has unusual relativistic effects
which are counterintuitive for electrons with speeds much slower
than the speed of light~\cite{graph1}. For example, it has been
recently shown~\cite{KP-graph} that graphene could be used for
experimentally testing the so-called Klein paradox~\cite{Klein}.
This quantum-mechanical effect of relativistic particles
penetrating through high and wide potential barriers can be
illustrated with massless Dirac fermions in graphene with a
potential barrier controlled by an applied voltage. A high
potential energy barrier in graphene, as was shown in
Ref.~\onlinecite{lens}, can also act as an unusual electron lens,
due to the negative refraction of electron waves at the edge of
the barrier, in analogy to the negative refraction of
3D~\cite{veselago} and 2D~\cite{kiv,kats} electromagnetic waves.

Our goal here is to show that, due to the Dirac-like Hamiltonian
of graphene with a potential energy barrier, quantum oscillations
similar to the Shubnikov-de-Haas effect can be observed
\emph{without an applied magnetic field}. Below we prove that the
density of electron states in a graphene sheet with a potential
barrier should display quantum oscillations if the strength of the
barrier (i.e., the product $V_0 D$ of the barrier's height $V_0$
and width $D$) exceeds some threshold value. In these
oscillations, the barrier strength $V_0 D$, that can be controlled
by a gate voltage, plays the same role as the external magnetic
field in the Shubnikov-de-Haas effect.

The quantum oscillations predicted here originate from a new type
of electron states in graphene. Contrary to non-relativistic
quantum mechanics, where localized states can only exist inside
quantum wells, we find that the electron states in graphene can be
localized \emph{within the barrier}. We show that the energy
$E(q_y)$ of the localized states (versus the wave vector component
$q_y$ along the barrier) becomes \emph{non-monotonic} if $V_0 D >
\pi \hbar v_F$ ($v_F$ is the Fermi velocity). This produces
singularities of the density of electron states for energies where
$dE/dq_y = 0$. When the magnitude and/or width of the barrier
changes, the locations of the singularities move and periodically
cross the Fermi level, generating quantum oscillations of both
thermodynamic and transport properties, e.g., of the conductance
in the $y$-direction (along the barrier).

{\it Electron states localized in a barrier.---} The tunnelling of
relativistic particles through a finite-width potential barrier
has recently been studied in Refs.~\onlinecite{win1,win,KP-graph}.
Here we consider another type of electron waves that propagate
\emph{strictly along} the barrier and \emph{damp away from it}.
Our analysis shows that a step-like barrier (i.e., a single edge
of an infinitely-wide barrier) does not support such electron
waves, which would be an analog to surface electromagnetic waves
(plasmon-polaritons) at the interface between two different media.
Therefore, even though the potential barrier in graphene could act
as an electron lens~\cite{lens}, it \emph{cannot} provide the
perfect lensing (i.e., subwavelength image reconstruction) that is
possible for Veselago's lens in optics~\cite{sciam}. However, as
we show in this section, the electron waves in graphene can be
localized inside \emph{a finite-width} potential barrier.

We consider electron states in graphene with a potential barrier
located in a single-layer graphene occupying the $xy$-plane (see
Fig.~1). For simplicity, we assume that the barrier $V(x)$ has
sharp edges,
\begin{equation}\label{1}
  V(x)= \left\{%
\begin{array}{ll}
    0, & \quad |x| > D/2, \\
    V_0, & \quad |x| < D/2. \\
\end{array}%
\right.
\end{equation}
Electrons in monolayer graphene obey the Dirac-like equation,
\begin{equation}\label{2}
\hat{H}\psi = i\hbar \frac{\partial \psi}{\partial t}, \quad
\hat{H}=-i\hbar \, v_F \, {\bf \sigma} \,\cdot\, {\bf \nabla}
+V(x)
\end{equation}
where $v_F$ is the Fermi velocity and ${\bf \sigma}=(\sigma_x,
\sigma_y)$ are Pauli matrices.
\begin{figure}[hbpt]
\vspace*{-2cm}
\includegraphics[width=8cm]{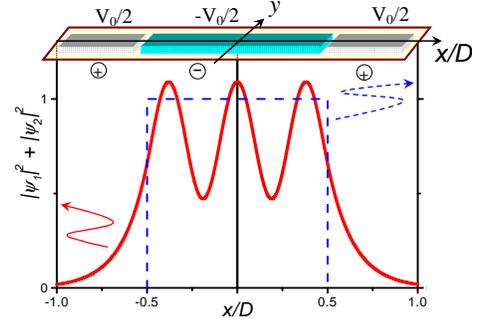}\vspace*{-6cm}
\caption{(Color online) (Top) Geometry of the problem. A graphene
sheet is placed under the voltage gates indicated by block
rectangles. (Bottom) Potential energy barrier $V(x)$ in graphene
(dashed blue line) and the probability distribution
$W(x)=|\psi_1(x)|^2+|\psi_2(x)|^2$ for the localized electron
state at $qd=3.5625$, $\epsilon D=0.003$, and ${\cal V}=9$ (red
solid line). }\label{f1}
\end{figure}
We seek stationary spinor solutions of the form,
\begin{equation}\label{3}
  \psi=\psi(x)\exp\left(-\frac{i}{\hbar}Et+iqy\right),
\end{equation}
with energy $E$ and momentum $\hbar q$ along the barrier. We focus
on the states with
\begin{equation}\label{4}
    |q|>|\epsilon| \equiv |E|/\hbar v_F.
\end{equation}
In this case, the electron waves satisfying Eq.~(\ref{2}) damp
away from the barrier, and the components $\psi_1$ and $\psi_2$ of
the Dirac spinor can be written in the form,
\begin{equation}\label{5}
  \psi_1(x)= \left\{%
\begin{array}{ll}
    a\exp(k_x x), & x<-D/2, \\
    b\exp(iq_x x)\\+c\exp(-iq_x x), & |x|<D/2, \\
    d\exp(-k_x x), & x>D/2, \\
\end{array}%
\right.
\end{equation}
\begin{equation}\label{6}
  \psi_2(x)= \left\{%
\begin{array}{ll}
    a\frac{i\epsilon}{(k_x+q)}\exp(k_x x), & x<-D/2, \\
    -b\exp(iq_x x+i\theta)\\+c\exp(-iq_x x-i\theta), & |x|<D/2, \\
   \frac{ -id\epsilon}{(k_x-q)}\exp(-k_x x), & x>D/2 \\
\end{array}%
\right.
\end{equation}
with real $k_x=(q^2-\epsilon ^2)^{1/2}$ and $q_x=[(\epsilon -
{\cal V}/D)^2-q^2]^{1/2}$. Here ${\cal V}=V_0 D/\hbar v_F$ is the
effective barrier strength and $\tan\theta = q/q_x$.

Matching the functions $\psi_1(x)$ and $\psi_2(x)$ at the points
$x=\pm D/2$, we obtain a set of four homogeneous algebraic
equations for the constants $a$, $b$, $c$, and $d$. Equating the
determinant of this set to zero, we obtain a dispersion relation
for the localizes electron states,
\begin{equation}\label{7}
   \tan(q_xD)=-\frac{k_xq_x}{({\cal V}/D-\epsilon)\epsilon + q^2}.
\end{equation}
Figure~1 illustrates the behavior of the probability distribution
$W(x)$ for a localized state. Note that $W(x)$ is an \emph{even}
function with \emph{continuous derivative} $W'(x)$, in spite of
the fact that each one of the functions $|\psi_1|^2$ and
$|\psi_1|^2$ are not even (the chirality of the Dirac spinors) and
both have discontinuous derivatives at the points $x=\pm D/2$.

Note that similar localized states can also be observed in a 2D
electron gas when a voltage is applied to produce a potential
well. For electrons with a quadratic dispersion law, this spectrum
is ${\cal E}_n=y_n^2+4Q^2-{\cal V}$, where ${\cal E}=mD^2E/2\hbar
^2$, ${\cal V}=mD^2V_0/2\hbar ^2$, $Q=qD$, $m$ is the electron
mass. Here $y_n$ is the $n$th root of the equation
$y\tan(y)=({\cal V}-y^2)^{1/2}$ for even states, and
$y\cot(y)=-({\cal V}-y^2)^{1/2}$ for odd states.

The spectrum Eq.~(\ref{7}) of localized states in graphene is
shown by the solid black curves in Fig.~2, for dimensionless
variables $Q=qD$ and ${\cal E} =\epsilon D$. This spectrum
consists of an infinite number of branches ${\cal E}_n(Q)$. Each
of these branches starts from the lines ${\cal E}=\pm |Q|$ (red
solid straight lines in Fig.~1b) at ${\cal E}={\cal
V}/2-\pi^2n^2/2{\cal V}$ and tends asymptotically to the lines
${\cal E}={\cal V}\pm Q$ (dashed red lines). Moreover, a
particular branch of the spectrum starts at the point ($Q=0$,
${\cal E}=0$) and also tends to the line ${\cal E}={\cal V} - Q$,
when increasing $Q$.
\begin{figure}[hbpt]
\vspace*{3cm}
\includegraphics[width=10cm]{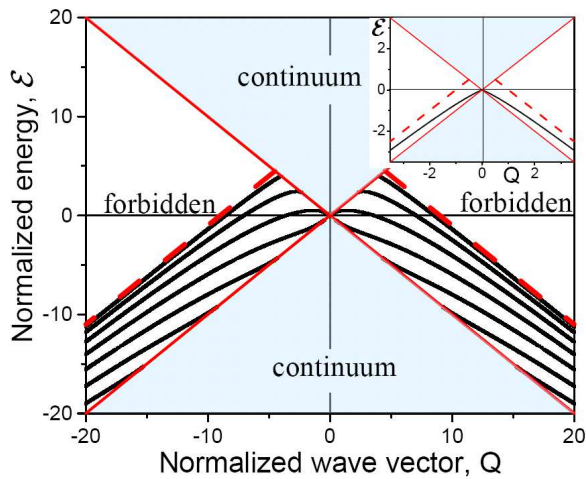}\vspace*{-3cm}
\caption{(Color online) Electron spectrum in graphene obtained for
${\cal V}=1$ (inset) and ${\cal V}=9$ (main panel). The sea of
delocalized states (continuum spectrum) is marked by the
light-purple regions. The branches of the spectrum for localized
states are shown by solid black curves between the straight solid
and dashed red lines. There are no states in the forbidden (white)
regions.}\label{f2}
\end{figure}

The behavior of different branches of the spectrum depends on the
barrier strength ${\cal V}$. If ${\cal V}<\pi/2$, all branches
satisfy ${\cal E}<0$ (see inset in Fig.~2). Localized states with
positive energies appear only for ${\cal V}>\pi/2$. When ${\cal
V}$ increases, new branches in the spectrum with positive energies
appear. When ${\cal V}$ is within the interval $(n+1/2)\pi <{\cal
V}< (n+3/2)\pi$, the number of branches with $E>0$ is $(n+1),\,
n=1,\,2,\,3,\,\dots$. We emphasize that each of the branches with
positive energy has a maximum ${\cal E}_n^{\rm max}$ at a certain
wave number $Q=Q_n^{\rm max}$. Near these points, the group
velocity of localized electron waves tends to zero. This effect is
similar to the stop-light phenomenon~\cite{stop} found in various
media, including superconductors~\cite{nature}.

Note that defect-induced localized electron states in graphene and
the enhancement of conductivity due to an increase of the electron
density of states localized near the graphene edges were recently
reported ~\cite{def,edge,mur2}. Contrary to these examples, the
electron states studied here are localized within the barrier and
also these \emph{are tunable}, i.e., the energy levels can be
shifted by changing the barrier strength (e.g., via tuning a gate
voltage).

{\it Density of localized electron states.---} To calculate the
density $N(E)$ of electron states, we use the general formula
$N(E)=\sum_\alpha \delta(E-E_\alpha)$, where $\alpha$ labels the
quantum state and $\delta (x)$ is Dirac's delta-function. Using
$\sum_\alpha \dots = 2L_xL_y(2\pi)^{-2}\int_{-\infty}^{\infty}dk_x
dk_y \dots$ for continuum spectrum, we derive
\begin{equation}\label{9}
N_{\rm cont}=N_0 |{\cal E}|, \quad N_0=\frac{L_xL_y}{\pi\hbar v_F
D},
\end{equation}
where $L_x$ and $L_y$ are the lengths of the graphene sheet in the
$x$ and $y$ directions, respectively. For localized states, we
obtain
\begin{equation}\label{10}
N_{\rm loc}({\cal E})=2N_0\frac{D}{L_x}\sum_n \left|\frac{d{\cal
E}_n(Q)}{dQ}\right|^{-1}_{{\cal E}_n(Q)={\cal E}}.
\end{equation}
Here $n$ runs over the number of positive roots of the equation
${\cal E}(Q)={\cal E}$.

The dimensionless density of states $N({\cal E})/N_0$ is shown in
Fig.~3. The localized electron states exhibit two types of
peculiarities. First, increasing ${\cal E}$, the jumps or steps
(each one of magnitude $2D/L_x$) in $N({\cal E})/N_0$ occur at the
points ${\cal E} = {\cal V}/2-\pi^2n^2/2{\cal V}$, where new
branches of the spectrum arise or disappear. More importantly,
singularities are observed when ${\cal E}={\cal E}_n^{\rm max}$,
where $|d{\cal E}_n/dQ|^{-1}$ in Eq.~(\ref{10}) diverges.
\begin{figure}[hbpt]
\vspace*{3cm}\includegraphics[width=10cm]{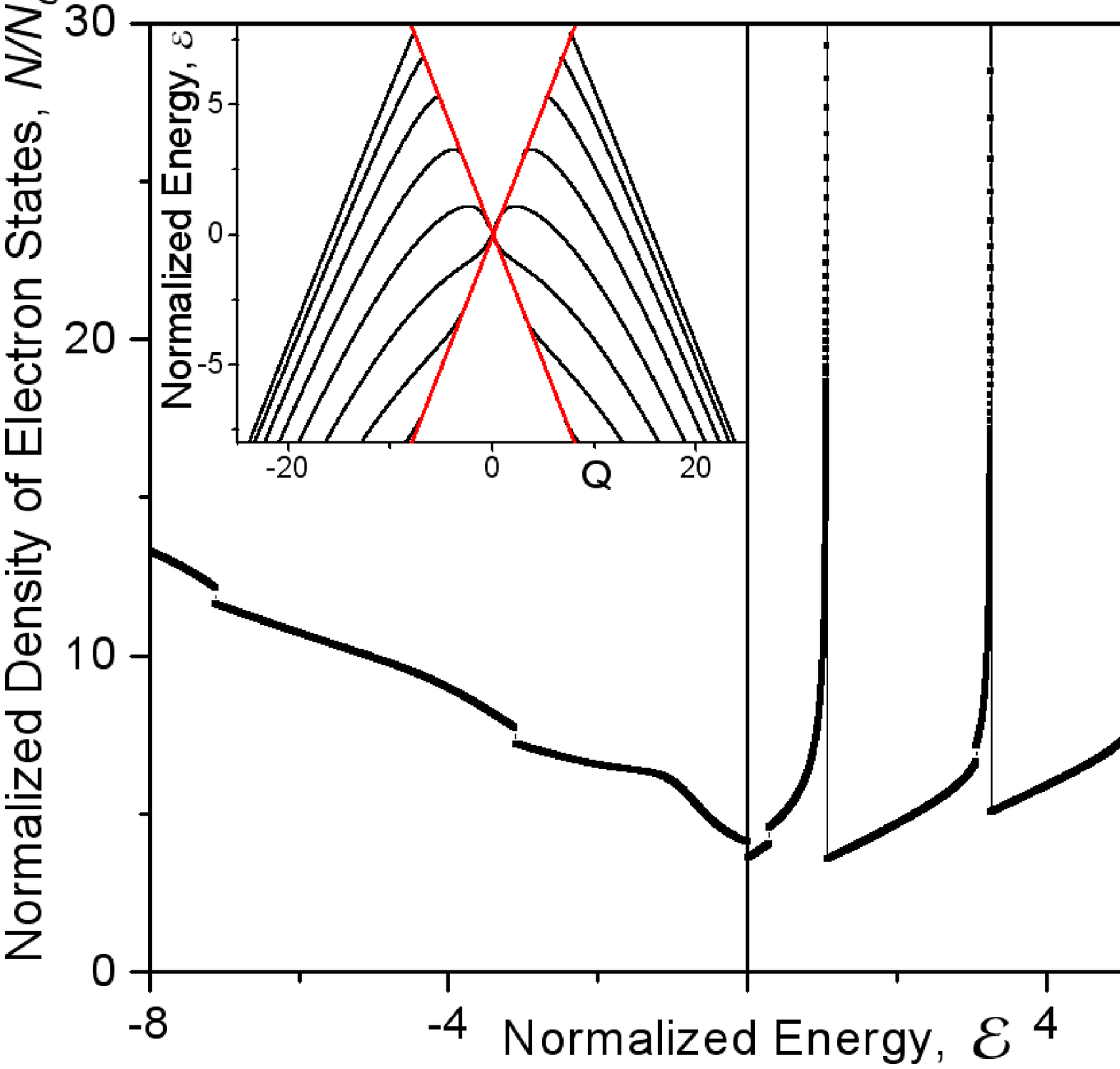}
\vspace*{-3cm}\caption{(Color online) Dimensionless density of
electron states $N({\cal E})/N_0$ in graphene with a potential
barrier, for $D/L_x=0.5$ and ${\cal V}=16$.}\label{f3}
\end{figure}

The locations of the singularities shift when changing the barrier
strength ${\cal V}$. Therefore, they periodically cross the Fermi
level ${\cal E}_F$. This produces quantum oscillations of the
density of states at the Fermi energy. They are seen in Fig.~4,
showing $N({\cal E}_F)/N_0$ versus the effective barrier strength
${\cal V}$.


\begin{figure}[hbpt]
\includegraphics[width=8cm]{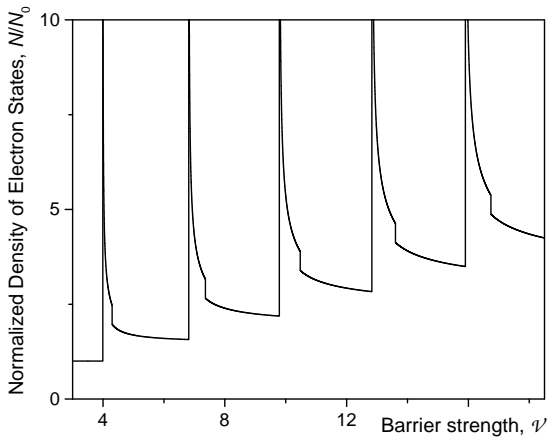}\vspace*{6cm}
\caption{Dimensionless density of electron states $N({\cal
N})/N_0$ at the Fermi level versus the effective strength ${\cal
V}$ of the potential barrier, for $D/L_x=0.5$ and ${\cal
E}_F=1$.}\label{f4}
\end{figure}

The periodic change in the number of electron states near the
Fermi level, increasing the barrier strength (e.g., by varying a
gate voltage), necessarily results in quantum oscillations of the
transport and thermodynamic properties of graphene. For example,
the conductance of graphene along the barrier qualitatively mimics
the quantum oscillations of the density of electron states.

Figure~4 shows the quantum oscillations of the density of electron
states in $e$-type graphene (with a positive Fermi energy). For
$p$-type graphene, with $E_F<0$, quantum oscillations of the
density of states at the Fermi level can also be observed, if the
(now opposite-bias) applied voltage forms a \emph{potential well
instead of a barrier}. Indeed, the Dirac equation (\ref{2}) is
invariant with respect to the transformation: $E \rightarrow -E,
\, V \rightarrow -V, \, x \rightarrow -x, \, y \rightarrow -y$.

In conclusion, we predict an unusual type of electron states in
graphene localized \emph{within a potential barrier}. For barriers
with sufficiently high magnitude and width, the density of
localized states has singularities. This feature of localized
states can result in quantum oscillations of the thermodynamic and
transport properties (e.g., the conductance along the barrier) of
graphene when changing the barrier strength (e.g., by varying a
gate voltage). These electric-field driven quantum oscillations
are similar to the Shubnikov-de-Haas oscillations of conductivity,
which are produced in standard metals when changing the external
magnetic field.

We acknowledge partial support from the NSA, LPS, ARO, NSF grant
No. EIA-0130383, JSPS-RFBR 06-02-91200, MEXT Grant-in-Aid No.
18740224, the EPSRC via No. EP/D072581/1, EP/F005482/1, ESF AQDJJ
network programme, and the JSPS CTC Program.

\end{document}